\def\bc{\begin{center}}                \def\ec{\end{center}}
\def\be{\begin{equation}}              \def\ee{\end{equation}}
\def\bear{\begin{eqnarray}}            \def\eear{\end{eqnarray}}
\def\la{\langle}  \def\ra{\rangle}     \def\l{\left}  \def\r{\right}
\def\i{\indent}                        \def\ni{\noindent}

\documentstyle[12pt]{article}
\begin{document}    
\begin{flushright} Preprint INRNE--TH--\,96/9\\
 quant-ph@9609001
\end{flushright}
\bigskip  \bigskip

\begin{center}
		{\large \bf ALGEBRAIC COHERENT STATES AND SQUEEZING}
\end{center}
\medskip
\centerline{\bf D.~A. Trifonov\footnote[0]{$^{a}$e-mail:
3fonov@phys.acad.bg}$^{a}$}
\centerline{Institute of Nuclear Research,}
\centerline{72 Tzarigradsko Chaussee,}
\centerline{1784 Sofia, Bulgaria}
\medskip \bigskip

\begin{center}

\begin{minipage}{13cm}
\centerline{\small {\bf Abstract}}
\medskip
{\small The eigenstates of general complex linear combination of
$SU(1,1)$ generators ($su^c(1,1)$ algebraic coherent states (ACS)) are
constructed and discussed.  It is shown that in the case of quadratic
boson representation ACS can  exhibit strong both linear and quadratic
amplitude squeezing. ACS for a given Lie group algebra contain the
corresponding Perelomov CS with maximal symmetry.
\vspace{0.5cm}

 }
\end{minipage}
\end{center}
\vspace{1cm}
\newpage
\baselineskip=18pt
\begin{center}
{\large {\bf I. Introduction}}
\end{center}
Canonical coherent and squeezed states (CCS and CSS) of quantum systems are
of considerable interest in many fields of physics, especially in quantum
optics (see the reviews [1] and references their in). CCS describe the laser
light and the CSS describe the squeezed light. These families of states have
been constructed [2,3] as eigenstates of photon (boson) destruction operator
$a$ (CCS [2]) and of complex linear combination $ua+va^\dagger$ (CSS or two
photon coherent states (CS)[3]).
The two quadrature components $q$ and $p$ of $a$, $[q,p]=i$ ($\hbar=1$), and
the unit operator $1$ close an algebra $wh$, known as Weyl-Heisenberg
algebra.  Thus the family of canonical SS consists of all eigenstates of
general complex combination of $1,\,q,\,p$, i.e. of general element of the
complexified algebra $wh^{c}$. One mode CSS (and only they) minimize the
Schr\"odinger uncertainty relation for $q$ and $p$ [5]. For the $n$ mode
field (or $n$ dimensional quantum mechanical system) the eigenstates of
complex linear combinations (of new lowering operators) $a^\prime_i =
u_{ij}a_j + v_{ij}a^\dagger_j$, $i=1,\,2\,\ldots$,
$[a^\prime_i,a^{\prime\dagger}_k] =
\delta_{ik} $, describe multimode squeezed light. It was shown [6] that
eigenstates of all $a^\prime_i$ (and only they) minimize the $n$ dimensional
Robertson uncertainty inequality [7]. The canonical transformations
$a_i,\,a^\dagger_i \rightarrow a_i^\prime, \,a^{\prime\dagger}_i$ are
automorphisms of $wh^c_n$ and they can diagonalize the $n$ mode uncertainty
matrix in any pure or mixed state [6]. Thus eigenstates of operators from
complex algebra $wh^c_n$ exhibit many interesting physical properties.

The aim of this paper is to construct the eigenstates of
general complex linear combination of generators of the group $SU(1,1)$,
to consider their squeezing properties and the possibility to
construct such states for other algebras.
 The continuous families of eigenstates  of complex Lie algebra operators
should be shortly referred to as {\it algebraic} CS (ACS).
ACS can be introduced for any Lie algebra for which at least one element has
normalized eigenstates. In particular ACS exist for any semisimple Lie
algebra and they contain as subsets the Klauder-Perelomov group related CS
with maximal symmetry [4,8]. The ACS are efficient to describe squeezing
(reduction) in fluctuation of observables related to hermitean operators
of the algebra, since by suitable restriction of parameters they could tend
to (or coincide with) the eigenstates of desired operator from the algebra.
They are suitable to describe transitions between eigenstates of different
operators of the algebra. If one succeeds to construct ACS one solves also
the spectral problem for the corresponding observables.

Eigenstate of the $SU(1,1)$ generator $K_- = K_1-iK_2$ in the bosonic
representation $K_-=a^2$ were constructed in [9] and called even and odd CS
.  The operators  $K_-$ in the  discrete series
$D^{(\pm)}(k)$, $k=\mp1/2,\,\mp1,\,\mp3/2,...$ have been diagonalized by
Barut and Girardello [10]. The next step is made in ref. [11], where
eigenstates of the complex combination $uK_-+vK_+$ ($K_{\pm} = K_1 \pm
iK_2$) for $D^{(-)}(k)$ are constructed and discussed.  In case of
Lipkin--Cohen representation (14) some particular linear combinations of
$a^2$, $a^{\dagger 2}$ and $a^\dagger a$ have been considered in
papers [13-16]. Here we construct the eigenstates of full linear complex
combination $\zeta^iK_i$ (summation over repeated indices) for square
integrable representations $D^{(-)}(k)$ and for the important in quantum
optics squared amplitude (bosonic) representation with Bargman indices
$k=1/4,\,3/4$ (eq. (14)).  We show that squared amplitude ACS exhibit many
nonclassical properties, such as strong squeezing of the quadratures of $a$
(amplitude squeezing) and/or of $a^2$ (squared amplitude squeezing) and
subpoissonian photon statistics.  The particular set of $su^c(1,1)$ CS which
are (normalized) eigenstates of the combination $ua^2+va^{\dagger 2}$,
$|v|<|u|$, constitute the full set of states which minimize the
Schr\"odinger uncertainty relation [17] for the two quadratures of $a^2$
[12].  States which minimize Schr\"odinger relation should be shortly called
Schr\"odinger intelligent states (Schr\"odinger IS, SIS).  Heisenberg IS are
SIS with vanishing covariance of the two
operators. Heisenberg IS for two generators of $SU(1,1)$ are considered also
in recent papers [14,15].
\bigskip

\begin{center}
{\large {\bf II. Algebraic CS }}
\end{center}
Let $L$ be real Lie algebra with basic elements $X_i$, $i=1,\,2,\,\dots,\,n$.
We assume that $X_i$ are realized as hermitean operators in Hilbert space
${\cal H}$ in order to represent physical observables. Then one can look for
the eigenstates $|z,\vec{\zeta}\ra$ of complex linear combinations
$\zeta^iX_i \equiv Z(\vec{\zeta})$, $\vec{\zeta} =
(\zeta^1,\,\zeta^2, \dots,\,\zeta^n),\quad z,\,\zeta^i\in C$,
\be    
Z(\vec{\zeta})|z,\vec{\zeta}\ra = z|z,\vec{\zeta}\ra.
\ee
The set of $Z(\vec{\zeta})$ span a Lie algebra,
called complexified $L$ and denoted  as $L^c$ [18].  Therefor the continuous
family of states $|z,\vec{\zeta}\ra$ (when exists and is at least dense in
${\cal H}$ [4]) should be called $L^c$ {\it algebraic} CS (ACS).

ACS can be constructed for many Lie algebras. They can be realized e.g.  for
any semisimple Lie algebra in the following way. Let $G_L$ be a Lie group with
parameters $\xi^i$ and $L$ as associated Lie algebra [18]. Consider the
operators $U(\vec{\xi})=\exp(i\xi^kX_k)$ which form a unitary representation
of $G_L$ in ${\cal H}$. Then we can take an eigenstate $|\psi_0\ra$ of the
operators $H_{\alpha}$ from Cartan subalgebra of $L$, $H_{\alpha}|\psi_0\ra
= h_\alpha|\psi_0\ra$ and construct the family of Klauder--Perelomov group
related CS $U(\vec{\xi})|\psi_0\ra$. Now we note
that $U(\vec{\xi})|\psi_0\ra$ are eigenstates (with the same eigenvalue
$h_\alpha$) of operators $H_\alpha^\prime\equiv U(\vec{\xi})H_\alpha
U^{-1}(\vec{\xi})$, which is easily seen (using BCH formula) to be a real
linear combination of $X_i$.  Thus group related CS
$U(\vec{\xi})|\psi_0\ra$ are particular case of ACS $|z,\vec{\zeta}\ra$.

The same group CS $U(\vec{\xi})|\psi_0\ra$ are also eigenstates of complex
combinations of $X_i$. Indeed, let $|\psi_0\ra$ be the highest (lowest)
weight vector. Then it is annihilated by Cartan raising (lowering) operators
$E_\alpha$ ($E_{-\alpha}$). Similarly the group CS $U(\vec{\xi})|\psi_0\ra$
are annihilated by the non hermitean operators $E^\prime_{\pm\alpha} \equiv
U(\vec{\xi})E_{\pm\alpha} U^{-1}(\vec{\xi})$, which clearly are complex
combinations of $X_i$. More general ACS we can get in the above form if in
$U(\vec{\xi})$ consider $\xi^i$ as complex parameters, but then we could get
also nonnormalized eigenvectors since $U(\vec{\xi})$ becomes nonunitary.
Most general normalized ACS $|{\rm acs}\ra$ one can get in this scheme if
one replaces $U(\vec{\xi})$ by unitary or at least isometric operator
$S(g_a)$ of the group $G_{A,L}$ of automorphisms of $L^c$:
\be    
|{\rm acs}\ra = S(g_a)|\psi_0\ra.
\ee
This construction is valid for any $L$, provided that $|\psi_0\ra$ is an
eigenstate of some operator from $L^c$.  The group $G_{A,L}$ is larger than
$G_L$,  $G_L \subset G_{A,L}$.  Therefor $L^c$ ACS contain the group related
CS for $G_L$. Known example of ACS is given by the squeezed CS[1], which are
ACS for the (non semisimple) nilpotent algebra $wh^c$.  The bose vacuum
$|0\ra$ is annihilated by $a\in wh^c$ and semidirect product group $WH\wedge
SU(1,1)$ is the group of automorphisms of $wh^c$. Then the $wh^c$ ACS (up to
a phase factor) take the known form [1],
\be    
|\alpha,\xi\ra \,=\, S(\xi)D(\alpha)|0\ra ,
\ee
where $D(\alpha)\in WH$ is the displacement, and $S(\xi)\in SU(1,1)$ is the
squeeze operator, $S(\xi) = \exp[(\xi a^{\dagger 2}-\xi^{*}a^2)/2]$.
Eigenstates of complex combination $ua+va^\dagger$ with $|u|^2-|v|^2=1$ have
been constructed and discussed as time evolved Glauber CS in refs. [19]
(they are the same as the two photon CS [3]).  In the next section we
construct the full set of ACS for the semisimple Lie algebra $su^c(1,1)$,
which has important quantum optics applications. Instead of looking for
explicit form of $S(g_a)$ and constructing orbits (2) here we solve directly
the eigenvalue problem (1) for $su^c(1,1)$.
\bigskip
\bc
{\large {\bf III. $su^c(1,1)$ CS }}
\ec
\medskip
 The generators $K_i$ of $SU(1,1)$ (the basic elements of the algebra
$su(1,1)$) satisfy the known commutation relations
\be   
[K_1,K_2]=-iK_3,\quad[K_2,K_3]=iK_1,\quad[K_3,K_1]=iK_2.
\ee
The complex linear combinations of these operators span the algebra
$su^c(1,1)$, which is isomorphic to $sl(2,C)$. The algebra $su(1,1)$ is
semisimple, so that according to the discussion in the preceding section,
the ACS here do exist.

To construct $su^c(1,1)$ CS we consider the eigenvalue problem for the
operators $Z \equiv \zeta^iK_i = uK_- + vK_+ +wK_3$,
\be   
(uK_- + vK_+ +wK_3)|\psi\ra = z|\psi\ra,
\ee
where $z$ and $\zeta_i$ are complex parameters and $u+v=\zeta_1$,
$i(v-u)=\zeta_2$, $w=\zeta_3$ . Such eigenstates should be denoted here as
$|z,u,v,w;k\ra$, $k$ being the Bargman index.  We shall solve the above
problem for the $D^{(-)}(k)$, $k=1/2,1,\dots\,$. It is then most suitable to
use the representation of Barut and Girardello CS (BG representation) [10]
for Hilbert space vectors and operators. In BG representation the group
generators $K_{\pm}$ and $K_3$ are differential operators,
\be   
K_+=\eta,\,\quad
K_-=2k\frac{d}{d\eta}+\eta\frac{d^2}{d\eta^2},\,\quad
K_3=k+\eta\frac{d}{d\eta},
\ee
where $\eta$ is a complex variable. We see that the eigenvalue equation (5)
becomes a second order linear differential equation for the eigenstates,
which in BG representation (in order to be normalized) should be entire
analytical functions $\Phi(\eta)$ of growth $(1,1)$ [10],
$$
\l(u\eta \frac{d^2}{d\eta^2} + (2ku+w\eta)\frac{d}{d\eta} +v\eta +
kw - z\r)\Phi_z(\eta) = 0. \eqno(5a)$$
  Note that
$\Phi_z(\eta;u,v,w)=\la k;\eta^*|z;u,v,w;k\ra$, where $|\eta;k\ra$ is BG CS
(eigenstate of $K_-$). Orthonormalized eigenstates $|m;k\ra$ of $K_3$ are
represented by $\eta^m\,[\Gamma(k)/(m!\Gamma(m+k))]^{1/2}$.

We shall consider first the case $u\ne0$ in (5a). By simple substitutions
the eq. (5a) is easily reduced to the Kummer equation [20], so that we have
the solution
\be   
\Phi_z(\eta;u,v,w)=N(z,u,v,w)\exp(c\eta)M(a,b,c_1\eta)
\ee
where $N(z,u,v,w)$ is a normalization constant, $M(a,b,\eta)$ is the Kummer
function (confluent hypergeometric function $_1F_1(a,b;\eta)$) [20] and
parameters $a,\,b,\,c$ and $c_1$ are
\bear  
a = k+\frac{z}{\sqrt{w^2-4uv}},\,\,\quad b=2k,\nonumber\\
c=-\frac{1}{2u}\l(w+\sqrt{w^2-4uv}\r),\quad c_1=\frac{1}{u}
\sqrt{w^2-4uv}.
\eear
$M(a,b,\eta)$ is an entire analytic function when $b\ne
-1,\,-2,...$, which holds in our case, where $b=2k>0$. It increases most
rapidly as $\exp(|\eta|)$, $|\eta|\rightarrow \infty,\, {\rm Re}\eta > 0$.
Therefor the solution (7) would have the required analyticity and
growth to represent normalized states $|z;u,v,w;k\ra$ when the
inequalities  $|c+c_1| < 1$ and $|c| = <1$ hold, i.e.
\be   
\frac{1}{2|u|}\l|w-\sqrt{w^2-4uv}\r| < 1, \qquad
\frac{1}{2|u|}\l|w+\sqrt{w^2-4uv}\r| < 1.
\ee
We note that if $a=-n$, $n=0,1,2,...$, i.e. the quantization condition
\be  
z=-(k+n)\sqrt{w^2-4uv}\equiv z_n,\quad n\,=\,0,1,2,\dots
\ee
is imposed the Kummer function becomes a polynomial [20]
of power $n$ and then only the second normalizability condition in (9) is
needed to ensure the required growth. In the special case of $l^2\equiv
w^2-4uv = 0$ both inequalities (9) are reduced to $|w/(2u)|<1$. Now we have
to take limit $l^2\rightarrow 0$ in solution (7): the Kummer function in
this limit is proportional to $_0F_1(2k,-z/u\eta)$.
Note that $l^2=(Z,Z)$, where $(,)$ is the Killing form [18].  When the
inequalities (9) are broken down the functions (7) still are solutions of
eq. (5a) and could be considered as non normalizable eigenstates.

Let us note some known particular cases of states (7). The BG CS $|z;k\ra$
[10] have been constructed as eigenstates of $K_-$. Therefor at $v=0=w$ our
states $|z,u,v,w;k\ra$ should recover the BG CS. And this is the case, as
one easily can check putting $v=0=w$ in $\Phi_z(\eta;u,v,w)$. Next,
according to the discussion in section II, we can recover the Perelomov CS
with maximal symmetry $|\tau;k\ra$ [8], $|\tau|<1$, in two natural ways
since these CS are eigenstates of $U(\vec{\xi})K_3U^{-1}(\vec{\xi})$ and are
annihilated by $U(\vec{\xi})K_{-}U^{-1}(\vec{\xi})$.
It was rather unexpected that Perelomov CS can be
reproduced in a third way, namely as a subset of $|z,u,v,w=0;k\ra$: if we put
\be   
w = 0, \quad {\rm and}\quad   z=-k\sqrt{-uv}
\ee
in $|z,u,v,w;k\ra$ then we get the CS $|\tau;k\ra$, $\tau=\sqrt{-v/u}$.
At $w=0$ the conditions (9) are reduced to $|v/u|<1$ so that the whole
family of Perelomov CS is recovered by the ACS $|z,u,v,w=0;k\ra\equiv
|z,u,v;k\ra$. It was shown [11] that these (and only these) states
$|z,u,v;k\ra$ minimize the Schr\"odinger uncertainty relation for $K_1$ and
$K_2$, i.e.  they are $K_1$-$K_2$ SIS [11] (or, in the terminology of ref.
[5], $K_1$-$K_2$ correlated states). Our large family $\{|z,u,v,w;k\ra\}$
recover all $K_i$-$K_j$ SIS by suitable restriction of parameters
$u,\,v,\,w$.
When the covariance of $K_i$ and $K_j$ vanishes (${\rm Im}(u^*v)=0$ in eqs.
(20)) SIS minimize Heisenberg relation. These particular $SU(1,1)$ SIS were
studied in the recent paper [15] using Perelomov CS representation.

The ACS $|z,u,v,w;k\ra$ can be easily expressed as series in terms of
orthonormalized states $|m;k\ra$. It is useful (with regards of their
possible generation) to represent them in a form, similar (but
not identical) to that of Perelomov CS,
\bear  
|z,u,v,w;k\ra = e^{\xi K_+ - \xi^*K_-}|\psi_0(u,v,w)\ra,\qquad\qquad\\
|\psi_0\ra = {\cal N}\,M\l(a,2k,\beta\tilde{K}\r)|0;k\ra;\quad
\tilde{K} = K_+ + 2c^*K_3 + 2(c^{*})^2K_-,
\eear
where $N$ is normalization constant, $a$ and $c$ is the same as
in eq. (8) and
$$\beta=\frac{1}{u}\l(1-\ln(1-|c|^2)\r)\sqrt{w^2-4uv},\quad |\xi| =
{\rm atanh}(|c|),\,\, {\arg}(\xi)={\arg}(c).$$
When $a=-n$ the "reference" state is a finite superposition of orthonormal
eigenstates of $K_3$, in particular, when $a=0$ (this is
$z=-k\sqrt{w^2-4uv}$) it is the "ground" state $|0;k\ra$ and formula (12)
becomes identical to that of Perelomov. Note, general $|z,u,v,w;k\ra$ can
not be put in Perelomov form. Moreover, one can prove that there is no
unitary operator $S$ which could relate $|z,u,v,w\!=\!0;k\ra$ to $|m;k\ra$
or to $|z;k\ra$ (but isometric $S$ do exists) [12]. Recall that these
two type of states have nontrivial stationary $su^c(1,1)$ subalgebra as it
is required in Perelomov construction [8].

The BG representation is valid for representations $D^{(\pm)}(k)$,
$k=1/2,1,\dots\,$ (here we consider $D^{(-)}(k)$). It is not valid for the
wider used (in quantum optics for example) Lipkin-Cohen representation,
characterized by $k=1/4,\,3/4$,
\be    
K_- = \frac{1}{2}a^2,\quad K_+=\frac{1}{2}a^{\dagger 2},\quad K_3 =
\frac{1}{2}(a^\dagger a +\frac{1}{2}).
\ee
Now it is most suitable to use the canonical CS representation [4] in which
$a=d/d\alpha,\,\,\,a^\dagger=\alpha$, $\alpha$ being complex variable, and
the normalized states are represented by entire analytical functions
$\Phi(\alpha;\psi)$ of growth $(1/2,2)$.
The eigenvalue eq. (5) for operators (14) can be again reduced to the Kummer
equation. Here we have two independent solutions of the form of entire
analytical functions of the required growth, therefor we write down the
general solution of the eigenvalue eq. (5) (for $u\ne 0$; the simpler case
$u=0$ should be solved afterwards)
$\Phi_{z}(\alpha) = C_{-}\,\Phi^{-}_z(\alpha)+C_{+}\,\Phi^{+}_z(\alpha)$.
The two independent solutions are
\be    
\Phi_z^{+}(\alpha;u,v,w) = N_{+}\,
\exp\l(c^{\prime}\alpha^2\r)\, M\l(a_{+}, \frac{1}{2}, c_2\alpha^2\r)
\equiv N_{+}\tilde{\Phi}_z^{+}(\alpha),
\ee
\be      
\Phi_z^{-}(\alpha;z,u,v) = \alpha N_{-}\,\exp\l(-c^{\prime}\alpha^2\r)\,
M\l(a_{-},{3\over 2},-c_2\alpha^2\r)\equiv
N_{-}\,\tilde{\Phi}_z^{-}(\alpha)
\ee
where $N_{\pm}$ are normalization constants and
\begin{eqnarray*}	%
 a_+ = \frac{1}{4}(1+ 2z/\sqrt{-uv^\prime}),\quad
 a_- = \frac{1}{4}(3 + 2z/\sqrt{-uv^\prime}),\\
 c^{\prime} = -\frac{1}{4u}\l(w+\sqrt{w^2-4uv}\r),\quad
 c_2= \frac{1}{2u}\sqrt{w^2-4uv},\quad v^\prime=-\frac{1}{4u}(w^2-4uv).
\end{eqnarray*}
The solutions (15) and
(16) represent normalizable even and odd states $|z;u,v,w;\pm\ra$ provided
$|c^{\prime}+c_2|<1/2$  and $|c^\prime|<1/2$ which result in the same
conditions (9) for $u,\,v,\,w$: both for $a_{\pm}\neq -n$ and the second one
only for $a_{\pm}= -n$. The last relation quantizes $z$ according to the
same formula (10) with $k=1/4$ and $k=3/4$.  The Kummer polynomials
$M(-n,1/2,\eta^2/2)$ and $M(-n,3/2,\eta^2/2)$ now are proportional to
Hermite polynomials $H_{2n}(\eta)$ and $(1/z)H_{2n+1}(\eta)$ respectively.
The corresponding discrete ACS are of the form $\exp(\xi K_+ -
\xi^{*}K_-)|\psi_0\ra$ with vector $|\psi_0\ra$ of the form
of finite superposition of Fock states $|n\ra$.
In the particular case of real $u,\,v$, $w=0$ and a
certain further restriction the squeezed Hermite polynomial states have been
constructed by Hillery et. al. [13]. Non normalizable eigenstates of
$(a+a^2\zeta)^2,\,\,|\zeta|=1$ were considered by W\"unshe[16].

In the same manner as for $k=1/2,1,\dots$ one can reveal the Perelomov CS
$|\tau;k=1/4,3/4\ra$ (which in quantum optics are known as squeezed vacuum
and one photon states) as subsets of ACS $|z,u,v,w;\pm\ra$. The other known
subset of $|z,u,v,w;\pm\ra$ are the Dodonov et. al. even and odd CS
$|\alpha_{\pm}\ra$ [8] (the Schr\"odinger cat states), which are correctly
recovered when $v=0=w$ in $|z,u,v,w;\pm\ra$:
$|z,1,0,0;\pm\ra=|(\sqrt{2z})_{\pm}\ra$.  When $w=0$ only we get the
eigenstates of $ua^2+va^{\dagger 2}$, which are constructed and discussed in
detail in  paper [12] as squared amplitude SIS.

To complete the set of $su^c(1,1)$ CS let us consider the case of $u=0$ in
the eigenvalue eq. (5).
In both BG and Glauber CS representations (for $k=1,\,1/2,\dots$
and for $k=1/4,\,3/4$ respectively) we have the first order differential
equation to solve. In both cases we get normalizable eigenstates provided
$|v/w|<1$. For the representations $k=1/4,\,3/4$ we have the solutions
\be     
\Phi_{z}(\alpha;v,w)\,=\,N\,e^{\tilde{c}\alpha^2}\alpha^b,
\ee
where $N$ is normalization constant, $\tilde{c}=-\frac{v}{2w}$,
$b=-\frac{1}{2}+\frac{2z}{w}$. In order at $v=0$ to get the eigenstates
$|n\ra \sim \alpha^n$ of $K_3=a^{\dagger}a/2 + 1/4$ we have to impose $b=n$,
i.e.  $z=w(n+2)/4\equiv z_n$. In Dirac notations we can represent solutions
(17) as squeezed binomial states
\be     
|n,v,w\ra=N\,S(\xi)\l(a^{\dagger} - \frac{v^*}{w^*}a\r)^n|0\ra,
\ee
where  $S(\xi)$ is the squeeze operator (see eq. (3)) and $\xi$ is defined
un terms of $v,\,w$ via $\tanh|\xi|=|v/w|,\,\,\arg\xi = \arg(-v/w)$. If in
(18) $v=0$ one gets the Fock states $|n\ra$.

In conclusion to this section let us note that in fact we have solved the
hole eigenvalue and eigenvector problem for general $su(1,1)$ hermitean
operators $X \equiv uK_- + u^*K_+ + wK_3 = X^{\dagger}$ in the above
representations. These operators can represent many physical observables, in
particular the Hamiltonians of some systems (e.g.  of the degenerate
parametric amplifier [1] and many other quadratic systems [8,21]). To reveal
the results one has simply to examine the normalizability inequalities (9)
for any specific combination $u,\,v=u^*$ and real $w$.  Since the
eigenstates of hermitean operators with different eigenvalues $x$ and
$x^\prime$ are orthogonal to each other we have established new
orthogonality relations between Kummer functions $M(a,2k,\eta) \equiv\,
_1F_1(a,2k,\eta)$ with different real $a$.
\bigskip

\begin{center}
{\large {\bf IV. The $su^c(1,1)$ CS and squeezing}}
\end{center}
\medskip
Algebraic CS are efficient in describing squeezing phenomena for the algebra
operators. The squeezing properties of the ACS stem from the observation,
that the (squared) variance $\Delta^2 X(\psi):=
\la\psi|X^2|\psi\ra - \la\psi|X|\psi\ra^2$ of an operator $X=X^\dagger$
vanishes iff $|\psi\ra$ is an eigenstate of $X$,
\be   
\Delta^2X(\psi) = 0\quad \Leftrightarrow \quad X|\psi\ra = x|\psi\ra.
\ee
If we know the eigenstates $|z,\vec{\zeta}\ra$ of combinations $\zeta^iX_i$
(i.e. the ACS (1) when $X_i$ close an algebra) then, when all but $\zeta^k$
parameters in $|z,\vec{\zeta}\ra$ vanish, $\zeta^i
\rightarrow 0$, $i\neq k$, the state $|z,\vec{\zeta}\ra$ is expected to tend
to the eigenstate of $X_k$. Then in virtue of (19) we would get $\Delta^2
X_k(\vec{\zeta}) \rightarrow 0$. If $X_k$ is with discrete spectrum then the
limit $\Delta^2 X_k = 0$ is expected to be reached.
In this section we shall examine for squeezing the constructed $su^c(1,1)$
CS.  The generators $K_{1,2}$ in the representation (14) appear as
quadrature components of squared photon annihilation operator $a^2$ and thus
here $K_{1,2}$ squeezing coincides with the "squared amplitude" squeezing
(quadratic squeezing) in quantum optics [13].  Since the linear amplitude
squeezing (i.e.  of $q$ and $p$) is important, we shall examine the new
states for it as well.

For any $su(1,1)$ hermitean representation the interest is in squeezing of
$K_1$ and $K_2$, since these operators have no normalizable
eigenstates [11] and therefor their variances never vanish exactly.
For this reason we shall examine ACS with $w=0$ for $K_{1,2}$ squeezing.
These are eigenstates of $K^{\prime} \equiv uK_- + vK_+$. Since the
normalizability condition now is $|v/u|<1$ it is convenient to set
$|u|^2-|v|^2=1$ which yields the commutator invariance
$[K^{\prime},K^{\prime \dagger}] = [K_-,K_+] = 2K_3$.  Then in any
eigenstate $|z,u,v,w\!=\!0;k\ra\equiv |z,u,v;k\ra$ of $K^{\prime}$ the three
second moments of $K_{1,2}$ are proportional to the mean of $K_3$,
\be      
\Delta^{2}K_1 = \frac{1}{2}|u-v|^2\la K_3\ra,\quad
\Delta^{2}K_2 = \frac{1}{2}|u+v|^2\la K_3\ra,\quad
\Delta K_1K_2 =  {\rm Im}(u^*v)\la K_3\ra.
\ee
The states $|z,u,v;k\ra$ tend to the eigenvectors of $K_1$ ($K_2$) when
$v\rightarrow u$ ($v\rightarrow -u$). Therefor we expect strong squeezing in
$K_{1,2}$ when $v\rightarrow \pm u$. We consider in greater detail the
squared amplitude representation. Using numerical integration in calculation
of the mean of $K_3$ we illustrate  the validity of the above statement on
the example of even ACS $|z,u,v;+\ra$ with parameters
$z\!=\!1,\,u\!=\!\sqrt{1+x^2},\,v \!=\! -x,\,\,x>0$ (see Fig.1). The
variance $\Delta K_2(x)$ is decreasing monotonically when $x$ is increasing
(that is $v\rightarrow -u$). For convenience we take the quadratures of
$a^2$ as  $X = (a^2+a^{\dagger 2})/\sqrt{2} = 2\sqrt{2}K_1$, $Y =
-i(a^2-a^{\dagger 2})/\sqrt{2} = 2\sqrt{2}K_2.$  Then in the ground state
$|0\ra$ of the oscillator (of the one mode electromagnetic field), $|0\ra =
|z\!=\!0,u\!=\!1,v\!=\!0;+\ra$, the variances of the above squared amplitude
quadratures $X,\,Y$ are both equal to $1$. Thus a state $|\psi\ra$ is
squared amplitude SS if $\Delta X(\psi)$ or $\Delta
Y(\psi)$ is less than $1$.
\bc \fbox{Fig. 1}  \ec
\bc {\small Fig.1. Squared variances of quadratures $p$ and $Y$ in states
$|z,u,v;+\rangle$,\\
$z=1$, $u=\sqrt{1\!+\!x^2}$,\, $v=-x<0$. Joint squeezing occurs
in $1.8<x<3.8$.}
\ec
As Fig.1 shows the algebraic CS
$|1,\sqrt{1+x^2},-x;+\ra$ are $Y$ squeezed when $x > 1.8$. We have to note
that states with strong squared amplitude squeezing have not been
constructed so far. The Heisenberg IS examined in ref. [13] exhibit relative
squeezing only, i.e.  $1<\Delta^2X_a < |\la[X,Y]\ra|/2$ ($X_a = X,\,\,Y$).

The ACS $|z,u,v;\pm\ra$ can exhibit also strong ordinary amplitude
squeezing (of the quadratures $q,\,p$ of $a$). The quadratures $q,\,p$ are
squeezed if their squared variance is less than $1/2$. In  $|z,u,v;\pm\ra$
we have
\be    
\Delta^2q = \frac{1}{2} + \la a^\dagger a\ra + {\rm Re}[(u-v)z^*],\quad
\Delta^2p = \frac{1}{2} + \la a^\dagger a\ra -{\rm Re}[(u-v)z^*].
\ee
On Fig.1 we show the plot of $\Delta^{2}p(x)$ for the same states
$|1,\sqrt{1+x^2},-x;+\ra$. $p$ squeezing occurs in the interval $0\leq x
\leq 3.8$. For larger $|z|$ the $p$ squeezing is stronger and occurs in
wider interval of $x$. It worth to underline that in the interval
$1.8\leq x\leq 3.8$ the states $|1,\sqrt{1+x^2},-x;+\ra$ exhibit $p$ and $Y$
squeezing simultaneously (joint squeezing). The states with opposite $z$,
$|-1,\sqrt{1+x^2},-x;+\ra$, are $q$ and $Y$ SS simultaneously in the same
interval $1.8\leq x\leq 3.8$. One can prove that a necessary
condition for $A,\,B$ joint squeezing is non positive (non
negative) definiteness of the commutator $i[A,B]$ [12].

Let us note that the eigenstates of $K^\prime$ and only they minimize [11]
the Schr\"odinger inequality [17] for $K_1$ and $K_2$:
$\Delta^2K_1\Delta^2K_2 - \Delta^2K_1K_2 \geq \la K_3\ra^2/2$. It is easy to
check that the three second moments (20) minimize this inequality
identically.  Therefor the states $|z,u,v;k\ra$ are $K_1$-$K_2$ SIS.  As we
have already seen for $k=1/4,\,3/4$ these states, $|z,u,v;\pm\ra$,  contain
as subsets the canonical squeezed vacuum and squeezed one photon states.
So we get that the squeezed vacuum states are very
symmetric -- they (and only they) minimize the Schr\"odinger relation for
both pairs $q,\,p$ and $X,\,Y$ and could be called double intelligent states
(IS). The squeezed one photon states are $X$-$Y$ IS only.

Quadratic squeezing occurs also in other ACS $|z,u,v,w;\pm\ra$ with $w\neq 0$.
For example light squeezing of $X$ and
$Y$ is found in the squeezed Schr\"odinger cats
\be   
S(\xi)|\alpha_+\ra = S(\xi)|z;+\ra \equiv |z,\xi;+\ra,
\ee
where $z=\alpha^2/2$ and $K_-|z;\pm\ra = z|z;\pm\ra$, $K_- = a^2/2$.
The identification with ACS is $|z,\xi;\pm\ra =
|z,u,v,w;\pm\ra$ where
$$u=\cosh^2r,\,\,\,v=\sinh^2r\,e^{2i\theta},\,\,\,w=\sinh(2r)e^{i\theta},\quad
\xi=r\,e^{i\theta}.$$
In these states
\bear 
\Delta^2\tilde{q}_{\pm}=\frac{1}{2}+\la a^\dagger a\ra\pm {\rm Re}\la a^2\ra,
\qquad\\
\Delta^2\tilde{X}_{\pm}=1+2\la a^\dagger a\ra+\la a^{\dagger 2}a^2\ra \pm
{\rm Re}\la a^4\ra - \la\tilde{X}_{\pm}\ra^2,
\eear
where $\tilde{q}_{+}=q,\,\,\tilde{q}_{-}=p,\quad \tilde{X}_{+}=X,\,\,
\tilde{X}_{-}=Y$. The means involved in above eqs. are
\be    
\la X\ra=\sqrt{2}{\rm Re}\la a^2\ra,\qquad
\la Y\ra = \sqrt{2}{\rm Im}\la a^2\ra,
\ee
\be   
\la a^\dagger a\ra=\sinh^2r+2|z|\sinh(2r)\cos(\theta-\phi)+\bar{n}\cosh(2r),
\ee
\be   
\la a^2\ra=\frac{1}{2}\sinh(2r)e^{i\theta}+2|z|\l(\cosh^2r\,e^{i\phi}
	+\sinh^2r\,e^{i(2\theta-\phi)}\r)+\bar{n}\sinh(2r)e^{i\theta},
\ee
\bear  
\la a^{\dagger 2}a^2\ra=\bar{n}\l(2\sinh^2(2r)+4\sinh^4r+2|z|\sinh(4r)
\cos(\theta-\phi)\r)\qquad\qquad\qquad\nonumber\\
	+ 4|z|^2\l(\sinh^2(2r)+\cosh^4r+\sinh^4r+(1/2)\sinh^2(2r)
	\cos(2\theta-2\phi)\r)\nonumber\\
	+2|z|\sinh(2r)\cos(\theta-\phi)\l(\cosh^2r +5\sinh^2r\r)+(1/4)
	\sinh^2(2r)+2\sinh^4r,
\eear
\bear  
\la a^4\ra=\bar{n}\l(3\sinh^2(2r)e^{2i\theta}+4|z|\sinh(2r)\l(\cosh^2r
e^{i(\theta+\phi)}+\sinh^2re^{i(3\theta-\phi)}\r)\r)\nonumber\\
	+4|z|^2\l((3/2)\sinh^2(2r)e^{2i\theta}
	+\sinh^4re^{2i(2\theta-\phi)}+\cosh^4re^{2i\phi}\r)\nonumber\\
	+6|z|\sinh(2r)\l(\cosh^2re^{i(\theta+\phi)}
	+\sinh^2re^{i(3\theta-\phi)}\r)+ (3/4)\sinh^2(2r)e^{2i\theta},
\eear
where $\bar{n}=\la+;z|a^\dagger a|z;+\ra$, $z=|z|e^{i\phi}$. On Fig.2 plots
of $2\Delta^2q(d)$ and $\Delta^2X(d)$ are shown for the states $|z,\xi;+\ra$
with $z=-d=-|z|,\,\, \xi=0.31$. In the latter states the variance of $X$ is
lightly squeezed for $0.1<d<0.31$, and the variance of $q$ is squeezed for
$0.17<d<0.51$.  In the interval $0.17<d<0.31$ both variances are squeezed
(joint $q$ and $X$ squeezing).  There are other states $|z,\xi\ra$ in which
the variance of $q$ or $p$ tend monotonically to zero (very strong linear
amplitude squeezing), e.g. $\Delta p \rightarrow 0$ in $|id,\xi=ir;+\ra$
when $r$ increases, $d$ being fixed.
\bc \fbox{Fig. 2} \ec
\bc {\small Fig.2. Squared variances of quadratures $q$ and $X$ in
states $|z,\xi;+\rangle$,\\
$\xi=0.31$, $z=-d$, $d>0$. Joint squeezing occurs in $0.17<d<0.31$.}
 \ec
The photon statistics in the above squeezed ACS is superpoissonian.
Subpoissonian statistics occurs in many of these states, e.g. in
$|z,u,v;+\ra$ with $z=-0.5-5i,\,v=-0.5,\,u=\sqrt{1.25}$ and $z=\pm 2.5,\,
u=\sqrt{1+x^2},\, v=x$, where $0<x<0.5$.  However these nonclassical states
are not squeezed.
\bigskip
\begin{center}
{\large {\bf IV. Concluding remarks }}
\end{center}
\medskip

We have shown that eigenstates of complex linear combinations of generators
of semisimple Lie group exist and contain the Perelomov CS with maximal
symmetry as particular case. It is natural to call such states {\it
algebraic} CS. Eigenstates of linear combination of any two particular
generators minimize the Schr\"odinger uncertainty relation. We noted that
these states are efficient in describing squeezing phenomena in quantum
physics
and demonstrated this on the example of $su^c(1,1)$ ACS. In the
latter case we have constructed the whole family $\{|z,u,v,w;k\ra\}$ of such
CS, solving the eigenvalue equations for the general element of $su^c(1,1)$
in irreps $D^{(-)}(k)$, $k=1/2,\,1,\dots$ and in the Lipkin--Cohen
representation (14), which is very important e.g. in quantum optics. In the
last case the even and odd ACS $|z,u,v,w;\pm\ra$ exhibit strong linear and
squared amplitude squeezing, even simultaneously.

The problem of physical realization of $su^c(1,1)$ states $|z,u,v,w;k\ra$ is
reduced to realization of Kummer function states, eq. (13), which for
$k=1/4,\,3/4$ are Hermite polynomial states. Polynomial states as a finite
superposition of Fock states in principle can be experimentally
constructed, as reported recently [22]. Then our discrete ACS can be
generated using these polynomial states as input in the degenerate amplifier
scheme.  Squeezed Schr\"odinger cat states, eq. (22), are an other subset of
ACS which can be generated in the same scheme because
CS $|\alpha_\pm\ra$ are available [23]. Since the field is better
determined in states with
joint linear and quadratic squeezing such states could be useful in
interferometric measurements [24].

After the first e-print submission my attention was  kindly brought  to
preprints \cite{Brif} where it was also noted that group related CS are
eigenstates of elements of complexified Lie algebra and
eigenstates of complex combinations of $SU(2)$ and of $SU(1,1)$ generators
were constructed using Glauber and Perelomov CS representations.

{\bf Acknowledgments}. The work is partially supported by Bulgarian Science
Foundation under Contracts \# F--559, F--644.
\bigskip
\begin{center}
{\large {\bf References}}
\end{center}
\medskip
\ni
[1] D.F. Walls, Nature (London) {\bf 306} (1983) 141;
		R. Loudon and P.L. Knight,\\[0pt]
	\i J. Mod. Opt. {\bf 34} (1987) 709; V.V. Dodonov and V.I. Man'ko,
		in Proc. Lebedev \\[0pt]
	\i Phys. Inst., v.183, (Nuova Science, Commack, N.Y., 1989);
		ibid. v.200 (1993).\\[0pt]
[2] R.J. Glauber, Phys. Rev. {\bf 131} (1963) 1726.\\[0pt]
[3] H. Yuen, Phys. Rev. A{\bf 13} (1976) 2226.\\[0pt]
[4] J.R. Klauder and B.S. Skagerstam. {\it Coherent States} (W.
		Scientific, Singapore,\\[0pt] \i 1985).\\[0pt]
[5] V.V.Dodonov, E. Kurmyshev and V.I. Man'ko, Phys. Lett.  A{\bf 76}
     	(1980) 150;\\[0pt]
	\i D.A. Trifonov, J. Math. Phys. {\bf 34} (1993) 100.\\[0pt]
[6] D.A. Trifonov, Preprint INRNE-TH-95/5. \\[0pt]
[7] H.R. Robertson,  Phys. Rev. {\bf 46} (1934) 794.\\[0pt]
[8] A.M. Perelomov, {\it Generalized Coherent States and Their Applications}
	(Springer,\\[0pt]
	\i Berlin, 1986; "Nauka", Moskva, 1987)\\[0pt]
[9] V.V. Dodonov, I.A. Malkin and V.I. Man'ko, Physica {\bf 72} (1974)
	597.\\[0pt]
[10] A.O. Barut and L. Girardello, Commun. Math. Phys. {\bf 21} (1971)
	41.\\[0pt]
[11] D.A. Trifonov, J. Math. Phys. {\bf 35}(5) (1994) 2297;
		Phys. Lett. A{\bf 187} (1994),\\[0pt] \i 284.\\[0pt]
[12] D.A. Trifonov, Preprint INRNE--TH--96/6 (to be submitted).\\[0pt]
[13] J.A. Bergou, M. Hillery  and  D. Yu, Phys.  Rev.  A{\bf 43}
	(1991) 515.\\[0pt]
[14] R.R. Puri and G.S. Agarwal, Phys. Rev. A{\bf
	53}(3) (1996) 1786; \\[0pt]
 \i A. Luis and J. Perina, Phys. Rev. A{\bf 53} (1996) 1886.  \\[0pt]
[15] C. Brif and Y. Ben-Aryeh, J. Phys. A{\bf 27} (1994)
	8185-8195. \\[0pt]
[16] A. W\"unsche, Acta Phys. Slovaca {\bf 45}(3) (1995) 413-424. \\[0pt]
[17] E. Schr\"odinger, in Sitz. Preuss. Acad. Wiss. (Phys.-Math. Klasse,
	p.296)\\[0pt]
  \i (Berlin, 1930); H.P. Robertson, Phys.Rev. {\bf 35}(5) (1930) 667.\\[0pt]
[18] A.O. Barut and Raszcka. {\it Theory of Group
	Representations and Applications}\\[0pt] \i (Polish Publishers, Warszawa,
	1977).	\\[0pt]
[19] I.A. Malkin, V.I.  Man'ko  and  D.A. Trifonov,  Phys.
	Lett. A{\bf 30} (1969) 413; \\[0pt]
	\i N. Cimento  A{\bf 4} (1971) 773; A. Holz, Lett. N. Cimento A{\bf 4}
	(1970) 1319. \\[0pt]
[20] {\it Handbook of mathematical  functions},  edited  by M. Abramowitz
	and I.A. Stegun\\[0pt]
	\i (National bureau of standards,  1964) (Russian translation, M.
	"Nauka", 1979).\\[0pt]
[21] I.A. Malkin, V.I.Man'ko, {\it Dynamical symmetries and Coherent
	States of}\\[0pt]
	\i{\it quantum systems} ("Nauka", Moscow, 1979)(in russian).\\[0pt]
[22] J. Janszky, P. Domokos, S. Szabo and P. Adam,
	Phys. Rev. A{\bf 51}(5) (1995) \\[0pt]\i4191.\\[0pt]
[23] L. Gilles, B. Garraway and P.L. Khight,
	Phys. Rev. A{\bf 49}(4) (1993) 2785; \\[0pt]
[24] N.A. Ansari, L. Di Fiore, M.A. Man'ko, V.I Man'ko, S. Solimeno\\[0pt]
    \i and F. Zaccaria, Phys. Rev. A{\bf 49} (1994) 2151.\\y0ptE
[25] C. Brif, quant-ph/9605006; C. Brif, submitted to Rep. Math. Phys.
\vspace{1cm}

\bc    F i g u r e  \hspace{5mm} c a p t i o n s    \ec
\bc {\small Fig.1. Squared variances of quadratures $p$ and $Y$ in states
$|z,u,v;+\rangle$,\\
$z=1$, $u=\sqrt{1\!+\!x^2}$,\, $v=-x<0$. Joint squeezing occurs
in $1.8<x<3.8$.}
\ec
\bc {\small Fig.2. Squared variances of quadratures $q$ and $X$ in
states $|z,\xi;+\rangle$,\\
$\xi=0.31$, $z=-d$, $d>0$. Joint squeezing occurs in $0.17<d<0.31$.}
 \ec
\end{document}